\documentclass{aa}
\usepackage{natbib,psfig}
\begin{document}

% \thesaurus{02.07.1; 03.13.8; 11.03.4;
% 11.05.1; 11.05.2; 11.06.1; 11.06.2; 11.19.6}

\title{Environmental effects on the galaxy luminosity function in the
cluster of galaxies Abell 496}

\author{ F. Durret\inst{1} \and C. Adami\inst{2} \and C. Lobo\inst{3,4}}

\institute{
Institut d'Astrophysique de Paris, CNRS, Universit\'e Pierre et Marie Curie,
98bis Bd Arago, 75014 Paris, France 
\and
LAM, Traverse du Siphon, 13012 Marseille, France
\and
Centro de Astrof\'\i sica da Universidade do Porto, Rua das Estrelas, 
4150-762 Porto, Portugal
\and
Departamento de Matem\'atica Aplicada, Faculdade de Ci\^encias,
Universidade do Porto, Rua do Campo Alegre, 687, 4169-007 Porto,
Portugal
}

\date{Accepted ???. Received ???; Draft printed: \today}

\authorrunning{Durret et al.}
\titlerunning{The galaxy luminosity function in the cluster of galaxies 
Abell 496}

\abstract{We have derived the galaxy luminosity function (GLF) in the
cluster of galaxies Abell 496 from a wide field image in the I band.
A single Schechter function reproduces quite well the GLF in the
17$\leq {\rm I_{AB}} \leq$22 ($-19.5\leq {\rm M_I} \leq -14.5$)
magnitude interval, and the power law index of this function is found
to be somewhat steeper in the outer regions than in the inner regions.
This result agrees with the idea that faint galaxies are more abundant
in the outer regions of clusters, while in the denser inner regions
they have partly been accreted by larger galaxies or have been dimmed
or even disrupted by tidal interactions.   }

\maketitle

\section{Introduction}\label{sec:intro}

Galaxy luminosity functions (hereafter GLF) are fundamental to analyse
the properties of galaxies in clusters. In a number of cases, it is
impossible to fit the entire GLF with a single Schechter function:
there appear to be two components in the GLF, one for the bright
galaxies - a gaussian distribution, and another for fainter galaxies -
a power law or a Schechter function (see e.g. Godwin \& Peach 1977,
Biviano et al. 1995, Durret et al. 1999a). This suggests that there
are at least two populations of galaxies in clusters, which do not
vary strongly from one cluster to another, since the dip between both
curves falls roughly at the same absolute magnitude in several
clusters (see e.g. Tab.~2 in Durret et al. 1999a).  Besides, at
faint magnitudes, the slope of the GLF can be steeper in the outskirts
of clusters i.e. in less dense environments, and flatter near the
cluster center (Lobo et al. 1997, Driver et al. 1998, Adami et
al. 1998, 2000, Andreon 2002, Beijersbergen et al. 2002). This can be
interpreted as due to the fact that in dense environments, small (and
faint) galaxies are more likely to be accreted by larger ones.
Moreover, they have also probably suffered repeated tidal interactions
on their way towards the cluster center, consequently being dimmed or
even disrupted (see e.g. Moore et al. 1996, Phillipps et al 1998,
Kajisawa et al. 2000).

We have performed a first analysis of the GLF of Abell 496 (Durret et
al.  2000) and intend to reobserve Abell 496 spectroscopically with
the VLT and VIRMOS; as a preparation, we asked R. Ibata and C. Pichon
to obtain for us a wide image of this cluster with the CFH12K camera
at CFHT. We present below our analysis of the GLF in different regions
at various distances from the cluster center.

\section{The data}

\subsection{Observations, reduction and detections}

\begin{figure}
\centering
\mbox{\psfig{figure=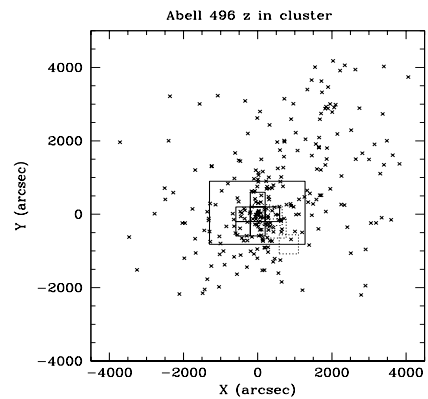,width=5.82cm,height=5.24cm}}
\caption[]{Positions of all the galaxies(relatively to the cD) with
redshifts in Abell 496 from the Durret et al. (1999b) catalogue
(crosses). The large rectangle shows the size of the present image;
the small squares forming a sort of cross in the center correspond to
the CCD catalogue by Slezak et al. (1999); the squares drawn with
dotted lines indicate the fields covered by Molinari et al. (1998).}
\label{fig:zin}
\end{figure}

Abell 496 (at redshift z=0.033, giving a distance modulus of 36.5,
assuming H$_0=50$ km~s$^{-1}$~Mpc$^{-1}$ and q$_0 = 0$, as also used
throughout this paper) was observed at the CFHT with the CFH12K camera
in the Mould I band on February 20, 2001. Two images of 5 minutes
exposure time each were obtained.  The CFH12K camera is made of 12
2K$\times$4K CCDs (hereafter, we will label the various CCDs from A to
L anticlockwise from the south east corner).  A global sketch of our
field, together with those previously observed by Molinari et
al. (1998) and by our team are shown in Fig. \ref{fig:zin},
superimposed on the positions of the galaxies with redshifts belonging
to the cluster (see the Durret et al. (1999b) catalogue).  The pixel
size of our image is 0.206 arcsec and the seeing 0.75 arcsec. The
interference fringes were corrected for and the photometrical
calibration was estimated from the observation of the Selected Area
101 in the I Kron-Cousins system (Landolt 1992), then converted to the
${\rm I_{AB}}$ system by ${\rm I_{AB}}$=I+0.456 (Fukugita et
al. 1995). All our I magnitudes will hereafter be ${\rm I_{AB}}$
magnitudes. The images were co-added and checked astrometrically at
the TERAPIX data processing center, leading to a final image of
12365$\times$8143 pixels, or 42.45$\times$27.96=1187 arcmin$^2$ in the
East-West and North-South directions respectively.

The sources were extracted using the SExtractor package (Bertin \&
Arnouts 1996). Saturated objects (SExtractor flag $\geq 4$) were
eliminated. The total number of objects thus eliminated was 154 in the
magnitude range 17$\leq {\rm I_{AB}} \leq$22 of interest here (see
below).  Their number varies from 0 to 19 objects from one CCD to
another, except for CCD L which has 69 (a first reason to discard this
CCD). So, CCD L apart, these numbers are at most 4\% of the total
number of galaxies used to derive the GLF in each CCD and therefore
eliminating them cannot strongly influence our results.

In order to avoid false detections at the edges of each of the 12
CCDs, the catalogue of detected objects was limited to an area
decreased by 15 pixels on all sides of each CCD. We thus obtained a
final catalogue of 37058 objects in a total area of 1123 arcmin$^2$
(or 33596 objects in 1031 arcmin$^2$ if CCD L is excluded).  This
catalogue will be made available in electronic form, with the
following Cols.: (1)~running number; (2) and (3) right ascension and
declination; (4) major axis $a$ in arcsec; (5) major axis position
angle; (6) ellipticity; (7)-(8) integrated elliptical Kron ${\rm
I_{AB}}$ magnitude and corresponding error; (9)-(12) aperture
magnitudes within 15, 10, 5 and 3.64 pixels respectively (3.64
pix=0.75 arcsec, the FWHM of the seeing).  All these parameters are
those estimated with SExtractor.  Since we have only one filter and a
rather short exposure time we cannot give any accurate morphological
information.  However, the combination of some of the information
provided in this catalogue, such as the total magnitude versus the
magnitude in an aperture having a diameter equal to the seeing FWHM,
can be used to obtain a first order morphological indication in the
form of a concentration parameter.

We have checked our photometry with data in the literature. First, we
identified 19 bright galaxies in common with the LEDA data base
(http://leda.univ-lyon1.fr). We find a mean value $<{\rm B_T - {\rm
I_{AB}}}>=1.89 $ ($\sigma$=0.36). Assuming ${\rm I_{AB}}$=I+0.456 this
gives $<{\rm B_T - I}>=2.35 $, in agreement with the value given for
elliptical galaxies by Fukugita et al. (1995): B-I=2.27.  Second, we
retrieved the Moretti et al. (1999) catalogue for Abell 496 (which is
broader and deeper than our previous R band catalogue) in the Simbad
data base. For 36 galaxies we obtain $<$r-${\rm I_{AB}}>$=0.70
(dispersion 0.53) and $<$i-${\rm I_{AB}}>=0.55$ (dispersion 0.55);
these values correspond to $<$r-I$>$=1.16 and $<{\rm i-I}>=1.01$ with
the above conversion, to be compared with the respective values of
1.04 and 0.75 given by Fukugita et al. (1995).  Therefore, the
agreement with the magnitudes of the Moretti catalogue is correct,
despite a possible shift by at most $\sim 0.25$ magnitudes. The
agreement is good with the LEDA catalogue.

\subsection{Completeness}

\begin{figure}
\centering
\mbox{\psfig{figure=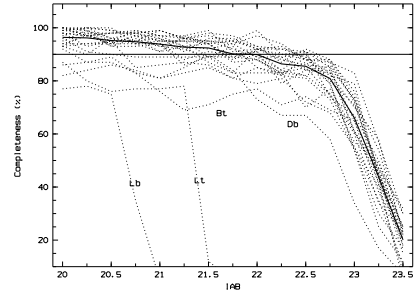,width=6cm,angle=-90}}
\caption[]{Variation of the completeness level as a function of I
magnitude for 24 different regions of the image (each CCD was split
horizontally into 2 equal sub-areas labelled t (top) and b (bottom)).
The thick curve shows the completeness averaged over all CCDs but
L. The thick horizontal line indicates the 90$\%$ completeness level.}
\label{fig:compl}
\end{figure}

We have carried out simulations to compute the completeness level. We
added artificial objects (similar to real objects) to our image and
measured the fraction of these objects recovered by the SExtractor
package as a function of magnitude and location in the image. For
this, we used a code created by J.M. Deltorn and already applied to
the CFDF survey (e.g. McCracken et al. 2001) to compute the star
detection completeness level. We modified this code in order to have a
more realistic representation for galaxies, and used a gaussian
profile with a FWHM of 3 times the mean seeing of our observation. The
results are given in Fig.~\ref{fig:compl}. This figure represents the
percentage of completeness level as a function of I magnitude for 24
different regions of the total image (each CCD was split horizontally
into 2 equal sub-areas). CCD L is significantly less complete than the
other CCDs, because of many dead columns, and was removed in the
following analyses. For the other CCDs, the mean completeness level is
close to 90$\%$ up to ${\rm I_{AB}}$=22, which will be taken as the
90\% completeness limit for our catalogue.  We also obtained another
estimation of our completeness level by comparing our observations
with those of the CFDF survey (e.g. McCracken et al. 2001), which used
the same instrument and filter, but with an exposure time of 5.5 hours
and a seeing of 1 arcsec, and was complete up to I=25.6. Their
completeness limit rescaled to our exposure time gives a completeness
level at I=21.8, in agreement with our simulations.  We will not
attempt to correct our counts for incompleteness at magnitudes fainter
than ${\rm I_{AB}}$=22, and we will hereafter limit our analysis to
${\rm I_{AB}} \leq$22.

In order to estimate the influence of ``crowding'' we computed the
number of galaxies susceptible to be masked by bright galaxies. For
18$\leq {\rm I_{AB}} \leq$22 there are 9726 galaxies in an area of
$1.007\times 10^8$ pixels$^2$ (entire field).  As a conservative
approach, we consider that galaxies brighter than ${\rm I_{AB}=18}$
and with surfaces larger than 600 pixels$^2$ can mask faint
galaxies. The total surface covered by these galaxies is 24130
pixels$^2$, leading to a number of galaxies in the 18$\leq {\rm
I_{AB}} \leq$22 magnitude interval that can be masked of the order of
a few. Therefore, the influence of crowding on our study appears to be
negligible.

\section{Estimating the background contamination}

Since the galaxy-star separation becomes difficult for magnitudes
${\rm I_{AB}}>20$, we decided to subtract the star and background galaxy
contaminations statistically.

\subsection{Star counts}

In order to subtract the stellar contribution from our Galaxy, we
produced a catalogue of stars using the Besan\c con model (Gazelle et
al. 1995) in the I Kron-Cousins band in the direction of Abell
496. The relation between the magnitudes measured with the two filters
is again: ${\rm I_{AB}}$=I$_{\rm Besancon}+0.456$. The uncertainty on
the star counts is smaller than 10\% for ${\rm I_{AB}} \leq$22
(A. Robin, private communication), and the contribution of stars
remains small in any case (less than one tenth of the total counts in
our image, all objects considered), so star counts cannot influence
our galaxy counts by more than a few percent in the magnitude interval
17$\leq {\rm I_{AB}} \leq22$. An alternative way to perform this
correction would have been to use the VIRMOS star counts, which have
the advantage of having been obtained with the same CFH12K camera and
filter, but they are not representative of the stellar counts in the
direction of Abell 496, due to their lower Galactic latitude.

\subsection{Field galaxy counts}

\begin{figure}
\centering
\mbox{\psfig{figure=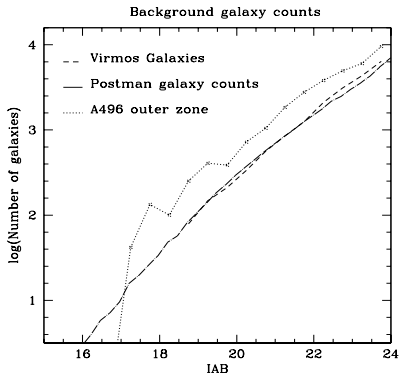,width=5.5cm,height=5cm}}
\caption[]{Field galaxy counts from the VIRMOS survey (dashed line)
and from Postman et al. (1998) (full line) a magnitude shift of
+0.456, normalized to the total area of 1123 arcmin$^2$ covered by our
catalogue. The galaxy counts derived from the outer regions of our
Abell 496 field as explained in the text (Sect.  3.2.) are also
shown.}
\label{fig:galaxies}
\end{figure}

An obvious way to correct for the background galaxy contribution would
be to extract from our image a region as far as possible from the
cluster center, subtract to it the star contribution and subtract the
resulting galaxy counts to our data. We have extracted such an ``outer
zone'' by putting together the data of the left half of CCD A and the
right halves of CCDs F and G, and derived the galaxy counts in this
zone.  As can be seen in Fig. \ref{fig:galaxies}, these counts are
higher than those issued from two independent field surveys (see
below), once all are normalized to the same surface area (the total
size of our image, i.e. 1123 arcmin$^2$), suggesting that the Abell
496 ``outer counts'' thus produced still contain a significant
fraction of cluster member galaxies, as confirmed by the positions of
cluster galaxies in Fig.  \ref{fig:zin}. So we would obviously
overestimate the background if we took it in this ``outer zone''.
Note that our image covers a total region of 2.404$\times$1.584
Mpc$^2$, while the r$_{200}$ radius calculated as in Carlberg et
al. (1997) with a velocity dispersion of 715 km/s (Durret et al. 2000)
is 2.5 Mpc, confirming that the cluster contribution in the outer
regions of our image is non negligible. Besides, the fact that
the last point of the counts in the outer zone (${\rm I_{AB}}$=23.75) does not
merge with any of the field survey background counts described below
adds still another reason to reject this type of background
subtraction.

We therefore decided to subtract the background contribution taken
from the VIRMOS survey galaxy counts (McCracken et al. in
preparation). However, the background galaxy contribution taken from
this survey shows an excess of objects at magnitudes brighter than
${\rm I_{AB}}$=18.5, and cannot be directly subtracted to our counts
either (see Fig. \ref{fig:galaxies}).  The comparison of the galaxy
counts by Postman et al. (1998) to the VIRMOS galaxy counts shows
similar slopes for ${\rm I_{AB}} \geq$18.5 (see
Fig. \ref{fig:galaxies}), with a magnitude shift due to the fact that
the VIRMOS counts are in ${\rm I_{AB}}$ magnitudes while the Postman
counts (which match well other counts such as those of Cabanac et
al. 2000) are in Cousins I magnitudes. Shifting the Postman counts by
0.456 magnitude (see Sect. 2.1) gives a good agreement between both
background counts for ${\rm I_{AB}} \geq$18.5.  We will therefore
subtract to our data the Postman galaxy counts shifted by 0.456
magnitude for 17$\leq {\rm I_{AB}} \leq$18.5 and the VIRMOS galaxy
counts for 18.5$< {\rm I_{AB}} \leq$22.
The star and background galaxy subtractions were performed in bins of
0.5 magnitude. The error bars that we indicate in the plots are simply
computed as the square root of the number of galaxies in the
corresponding magnitude bin.

\section{Fitting method and results}

\subsection{Fitting method}

Schechter function fits were performed using an IDL code based on the
$curvefit$ function, which uses a gradient-expansion algorithm to
compute a non-linear least squares fit to a given function; this
routine gives the best fit parameters and respective errors of the
Schechter function:
 $$ \Phi(M) dM = K \Phi^* 10^{0.4(M^*-M)(\alpha +1)}\exp (-10^{0.4(M^*-M)})$$
\noindent
where $\Phi^*$ is the normalisation, $M^*$ the characteristic apparent
magnitude, $\alpha$ the slope of the faint end of the luminosity function 
and $M$ the apparent magnitude of a given galaxy in the I band.

\subsection{Overall galaxy luminosity function}

\begin{figure}
\centering
\mbox{\psfig{figure=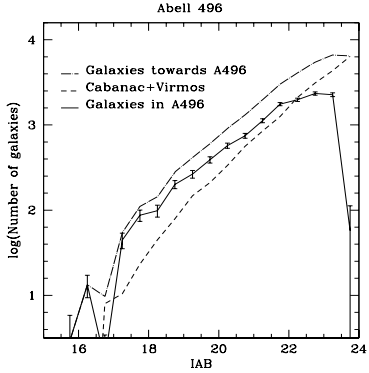,width=5.5cm,height=5cm}}
\caption[]{Galaxy counts in the direction of Abell 496 (dot-dashed
line), background galaxy counts -- see Sect. 3.2 (dashed line), and final galaxy
luminosity function for Abell 496 after background subtraction (full
line) with its error bars.}
\label{fig:fdlall}
\end{figure}

\begin{figure*}
\centering
\mbox{\psfig{figure=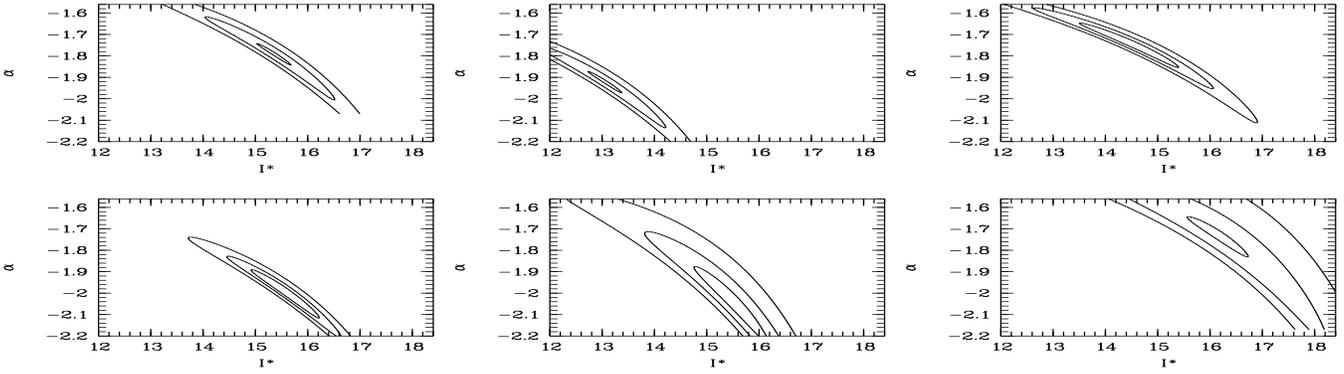,width=18.0cm,height=5cm,clip=yes}}
\caption[]{Correlation between the two parameters M$^*$ and $\alpha$
of the Schechter function for six regions: all the cluster, ABK, CDIJ
(top row), EFGH, F and C (bottom row). The isocontours are 1, 2 and
3$\sigma$ respectively.}
\label{fig:correlations}
\end{figure*}

The resulting GLF for the entire field (after eliminating CCD L, the
top half of CCD B: Bt and the bottom half of CCD D: Db) is shown in
Fig.~\ref{fig:fdlall}.  A single Schechter function was fit in the
same magnitude interval that we have been using (17$\leq {\rm I_{AB}}
\leq$22, $-19.5\leq {\rm M_I} \leq -14.5$), giving a power law index
slope $\alpha = -1.79 \pm 0.01$. Results are given in Table~\ref{tab:alpha}
for all regions that we explore. For each
case, we indicate the reduced $\chi ^2$; all values are about 1 or
lower, indicating that the fits are correct.  Since $M^*$ is brighter
than the lower limit of the magnitude interval considered here, it is
not well constrained and we will therefore focus our discussion on the
values of the slope $\alpha$ only. The correlation between $\alpha$
and $M^*$ is shown as confidence ellipses in
Fig. \ref{fig:correlations} for various regions.
The ellipses confirm that the error bars that we give on $\alpha$ in
Table~\ref{tab:alpha} are realistic. Note that the error bars in
Table~1 are 1$\sigma$ error bars and correspond to the innermost
ellipses in Fig.~\ref{fig:correlations}.  We made tests on region CDIJ
(see below), to see how an underestimate of the error bars could modify
the slope of the GLF. For this, we multiplied the error bars by
factors of 2 and 10 and found that $\alpha$ remains unchanged (even
though our fitting procedure takes the error bars into account), while
the uncertainty on $\alpha$ increases to $\pm 0.05$ and $\pm 0.1$
respectively, instead of the previous value of $\pm 0.03$. In order
for our results to lose significance, we would have to multiply the
error bars by 10, which seems an unrealistically large number (as seen
for example from the scatter in the Metcalfe et al. Fig. 13). A factor
of 2 seems much more probable, and in this case the difference in
slopes between CDIJ and other regions remains significant.

\begin{table}
\caption{Schechter law parameters for the various regions.}
\begin{tabular}{lrrrr}
\hline
Region & Nb. & Area &$\alpha$~~~~ & $\chi ^2_{red}$ \\
       & gal.  & (arcmin$^2$) & & \\
\hline
All$^*$ & 4052 & 1031 &	$-1.79\pm0.01$ & 7.26/7 \\
ABK     & 1217 &  281 &	$-1.93\pm0.02$ & 4.14/7 \\
CDIJ    & 1542 &  378 &	$-1.75\pm0.03$ & 1.50/7 \\
EFGH    & 1652 &  372 &	$-1.98\pm0.03$ & 1.80/7 \\
CenL    &  426 &   94 &	$-1.82\pm0.04$ & 1.80/7 \\
CenM    &  292 &   53 &	$-1.79\pm0.07$ & 0.48/7 \\
CenS    &  129 &   24 &	$-1.60\pm0.25$ & 1.50/7 \\
A       &  409 &   93 &	$-2.05\pm0.05$ & 3.60/7 \\
B       &  402 &   94 &	$-1.73\pm0.04$ & 3.18/7 \\
C       &  312 &   95 &	$-1.73\pm0.05$ & 3.60/7 \\
D       &  353 &   94 &	$-1.87\pm0.04$ & 3.36/7 \\
E       &  364 &   94 &	$-1.68\pm0.13$ & 1.98/7 \\
F       &  460 &   92 &	$-2.03\pm0.05$ & 2.34/7 \\
G       &  297 &   92 &	$-1.87\pm0.11$ & 2.76/7 \\
H       &  534 &   94 &	$-2.03\pm0.05$ & 4.08/7 \\
I       &  374 &   94 &	$-1.60\pm0.10$ & 0.66/7 \\
J       &  505 &   94 &	$-1.78\pm0.08$ & 2.46/7 \\
K       &  407 &   94 &	$-1.94\pm0.05$ & 4.08/7 \\
\hline
\end{tabular}
\label{tab:alpha}

$^*$ All but L,Bt,Db
\end{table}

\subsection{Galaxy luminosity function in three large regions}

\begin{figure}
\centering \mbox{\psfig{figure=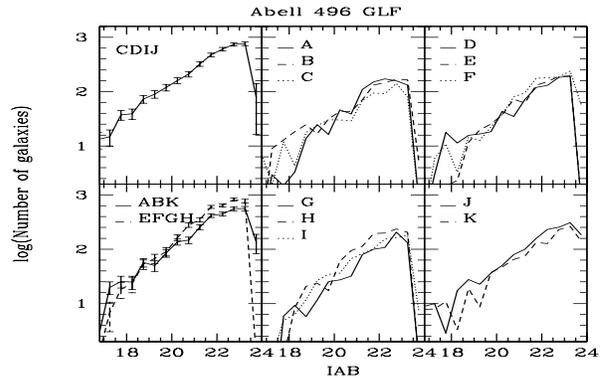,width=8.5cm,height=5.0cm}}
\caption[]{Galaxy luminosity function in various regions of Abell 496
(see text and Table~1). }
\label{fig:fdl6reg}
\end{figure}

\begin{figure}
\centering
\mbox{\psfig{figure=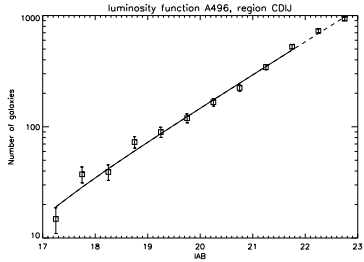,width=5cm}}
\caption[]{GLF in the central region (CDIJ) of Abell 496 with the best fit
Schechter function superimposed.}
\label{fig:schechter}
\end{figure}

We then divided the cluster into three regions of roughly equal
surface, CDIJ surrounding the cluster center (CCDs C, D, I and J), ABK
(CCDs A, B, K) towards the East and EFGH (CCDs E, F, G, H) towards the
West.  The GLFs in these three regions are shown in
Fig.~\ref{fig:fdl6reg}. The fit of the GLF by a Schechter function in
region CDIJ is displayed in Fig.~\ref{fig:schechter} (the fit was done
in the interval 17$\leq {\rm I_{AB}} \leq$22 even though the figure shows the
GLF in a larger range of magnitudes, for which we also extrapolated
this best fit).

The slope of the GLF is found to be flatter in region CDIJ, the
Schechter law slopes being $\alpha = -1.75 \pm 0.03$, $-1.93 \pm
0.02$, and $-1.98 \pm 0.03$, for regions CDIJ, ABK and EFGH
respectively (see Table \ref{tab:alpha}).

\subsection{Mapping the parameters of the galaxy luminosity function}
\label{sec:map}

The GLFs in the 11 CCDs (L excluded) are shown in
Fig.~\ref{fig:fdl6reg} and Table~\ref{tab:alpha}. Here also, the
Schechter function slope may be steeper in the outer regions of the
cluster, but better statistics are obviously required.

Finally, we selected three rectangular concentric regions of different
sizes centered on the intersection of CCDs C, D, I and J (or, roughly
the position of the cD cluster galaxy). CenL (Large), is of the size
of a CCD and covers one quarter of CCDs C, D, I and J.  CenM (Medium),
covers 9/16 of the area of CenL.  CenS (Small), covers 1/4 of CenL.
The Schechter fits of the luminosity functions give slopes
$\alpha=-1.82 \pm 0.04,-1.79\pm 0.07$ and $-1.60\pm 0.25$ for CenL,
CenM and CenS respectively, consistent with that in the CDIJ area
within error bars. Region CenS may exhibit a flatter slope, but this
needs confirmation since the uncertainty is very large.

\section{Discussion and conclusions}

We have derived the GLF in various regions of Abell 496.  The slope of
the Schechter function fit is always found to be steep (between
$-1.60$ and $-2.05$). Since such a steep slope could be due to several
artefacts, we will discuss the validity of our results. First, the
background counts could have been underestimated.  However, the good
agreement of the various background counts (VIRMOS, Postman, Cabanac),
and the fact that the subtraction is mainly that of the VIRMOS counts
(in the interval 18.5$< {\rm I_{AB}} \leq$22), made with the same
instrument, filter and magnitude system as ours, tends to suggest that
this is not the case. Second, the number of faint galaxies may have
been overestimated; for example, we may have confused globular
clusters with galaxies at faint magnitudes, as explained in detail by
Andreon \& Cuillandre (2002). However, we limit our sample to ${\rm
I_{AB}}$=22, where such effects should not be too strong. Third, our
${\rm I_{AB}}$ magnitudes may be too bright by 0.25 magnitude, as
suggested by the difference with the Moretti et al. data (see
Sect. 2.1). We tried to fit the GLF in several regions after shifting
the ${\rm I_{AB}}$ magnitudes by 0.25 (before subtracting the
background) and find slopes $\alpha= -1.68\pm0.05, -1.85\pm 0.03$ and
$-1.94\pm 0.04$ for regions CDIJ, ABK and EFGH respectively, instead
of the previous values of $-1.75, -1.93$ and $-1.98$. Therefore,
although the values change a little, the slope remains flatter in
CDIJ.

We then used our ${\rm I_{AB}}$ catalogue limited to the region in
common with Molinari et al. (1998) and made a Schechter fit as
described above. The Molinari et al. zone partially covers our CCDs
J,I,C,D,E and F, with a main concentration towards CCD E. A Schechter
fit in the same magnitude interval gives a slope $\alpha=-1.71 \pm
0.06$, close to the value of $-1.68\pm 0.13$ found in CCD E. A shift
of ${\rm I_{AB}}$ by 0.25 magnitude as above gives $\alpha=-1.68\pm
0.09$, in perfect agreement with Molinari.  Fits in broader magnitude
intervals give: $\alpha=-1.66\pm 0.05, -1.64\pm 0.04$ and $-1.57\pm
0.03$ for the magnitude intervals 17$\leq {\rm I_{AB}} \leq$22.5,
17$\leq {\rm I_{AB}} \leq$23, and 17$\leq {\rm I_{AB}} \leq$23.5
respectively. Molinari et al. give a slope $\alpha=-1.49\pm 0.04$ in
the i band, but mention that a magnitude correction allows them to
reach a slope as steep as $\alpha=-2.0$. We therefore believe that our
results are consistent with theirs.  Note that such a slope is not
much steeper than found e.g. in Coma (Lobo et al. 1997) or in Abell
665 (De Propris et al. 1995). This could indicate an excess of faint
red galaxies, but to ascertain this hypothesis it would be necessary
to derive the GLF in Abell 496 in other filters from samples of
comparable quality (covered area and depth).  As still another test on
the robustness of our results, we reanalyzed the GLF in the CDIJ
region. For this, we reduced the number counts by 10, 20, 30 40 and
50\% for ${\rm I_{AB}} >$20 (below ${\rm I_{AB}}$=20 the counts
remained unchanged) and made fits of these new GLFs.  Results are
given in Table 2. They show that the slope changes strongly only if
the counts are reduced by at least 30\%, an unrealistic number.
Besides, in order to account for the difference in slope of 0.2 that
we observe for example between regions CDIJ and EFGH, we would need to
make an unrealistically large error of 50\% on the counts.  We are
therefore confident that both the absolute values of $\alpha$ and
their variations from one zone to another are robust.

Our second result is that the slope of the Schechter function fit
tends to be steeper in the outer regions of the cluster, as already
observed in other clusters (see references in
Sect.~\ref{sec:intro}). Such a variation of $\alpha$ can be
interpreted as due to the fact that faint galaxies are accreted by
larger ones preferentially in the inner parts of clusters, where the
galaxy density is higher, therefore inducing a lack of faint galaxies
and a flattening of the GLF in the inner regions.  Moreover, galaxies
are likely to have suffered repeated tidal interactions on their way
towards the cluster center, consequently being dimmed or even
disrupted in a scenario of galaxy harassment (Moore et al. 1996).

The next step is obviously to confirm these results through deep
multiband imaging and/or spectroscopy that would make the background
subtraction more secure and would allow us to compare the GLFs in
various filters.

\begin{acknowledgements}

We are indebted to R. Ibata and C. Pichon for taking this image for
us, correcting it for interference fringes and calculating the
photometrical zero point.  We are grateful to the TERAPIX data center
for help in data reduction, in particular to M. Dantel-Fort for
reducing the data, co-adding the images and checking the astrometry,
and to E. Bertin for his help in using the SExtractor package.  We
thank O. LeF\`evre, S. Foucaud and all the VIRMOS team for allowing us
to use the VIRMOS background counts and C. Savine for help. Finally,
we are grateful to the referee, Stefano Andreon, for many helpful 
suggestions.  CL and FD acknowledge support from ESO/PRO/15130/1999
and PNC, CNRS-INSU.

\end{acknowledgements}

\begin{table}
\caption{Schechter law slope when the counts are reduced.}
\begin{tabular}{rr}
\hline
Reduction of & New values \\
counts       & of $\alpha$~~~~ \\
\hline
0\%  & $-1.75 \pm 0.03$\\
10\% & $-1.72 \pm 0.04$\\
20\% & $-1.68 \pm 0.03$\\
30\% & $-1.61 \pm 0.03$\\
40\% & $-1.59 \pm 0.02$\\
50\% & $-1.48 \pm 0.04$ \\ 
\hline
\end{tabular}
\label{tab:alphaerr}
\end{table}

\end{document}